\documentstyle[w-edbkps,psfig]{w-edbk}
\begin{document}

\tableofcontents

\title[Quantum-optical states in finite-dimensional Hilbert space. II. State
generation]{}

\markboth{}{Quantum-optical states in FD Hilbert space. II. State
generation}


\begin{center}

{\bf  QUANTUM-OPTICAL STATES IN\\ FINITE-DIMENSIONAL HILBERT
SPACE.\\ II.~STATE GENERATION}\footnote{Published in: {\em Modern
Nonlinear Optics}, Part 1, Second Edition, Advances in Chemical
Physics, Vol. 119, Edited by Myron W. Evans, Series Editors I.
Prigogine and Stuart A. Rice, 2001, John Wiley \& Sons, New York,
pp. 195--213.}

\vspace{4mm}%
WIES\L{}AW LEO\'NSKI

\vspace{3mm}%
{\em Nonlinear Optics Division, Institute of Physics, Adam
Mickiewicz University, Pozna\'n, Poland}

\vspace{4mm}%
ADAM MIRANOWICZ

\vspace{2mm}%
{\em CREST Research Team for Interacting Carrier Electronics,
School of Advanced Sciences, The Graduate University for Advanced
Studies (SOKEN), Hayama, Kanagawa, Japan and Nonlinear Optics
Division, Institute of Physics,\\ Adam Mickiewicz University,
Pozna\'n, Poland}

\end{center}

\section{~~~ I.~ Introduction}
\label{sectintr} \inxx{finite-dimensional Hilbert space}

As it was mentioned in the first part of this study \cite{MLI00},
the {\em finite-dimensional} (FD) quantum-optical states have been
a subject of numerous papers. For instance, various kinds of FD
coherent states \cite{BWKL92}--\cite{RR98},
FD Schr\"odinger cats \cite{MOB97,RR98,ZK94}, FD displaced number
states \cite{MOB97}, FD phase states \cite{PB89}, FD squeezed
states \cite{WKBB87,MLT98} were studied by many authors. In this
chapter we concentrate on some schemes of generation of the FD
quantum-optical states. These states can be produced as a finite
superposition of $n$-photon Fock states. As a consequence, the
problem of generation of FD states can be reduced to the choice
of the mechanism of $n$-photon Fock state generation. For
instance, Fock states can be achieved in the systems with
externally driven cavity filled with the Kerr media
\cite{K90}--\cite{LDT97}. Moreover, they can be produced in the
cavities using micromaser trapped states \cite{KSWW89}. Another
way to obtain Fock states is that proposed by D'Ariano {\em et.
al.} \cite{AMPS00} based on the {\em optical Fock-state
synthesizer}, in which the conditional measurements have been
performed for the interferometer containing Kerr medium. The
cavities with moving mirror \cite{BJK97} can also be utilized for
the FD state generation. Recently, several schemes for the
optical-state truncation ({\em quantum scissors}), by which FD
quantum-optical states can be produced via teleportation, have
been analyzed \cite{PPB98,truncation}. Various other methods for
preparation of Fock states \cite{Fock} and their arbitrary
superpositions \cite{state1} have been developed (see also Ref.
\cite{state2}).

However, we shall concentrate here on the generation methods in
which we are able to get directly the FD quantum state desired.
Namely, we shall describe the models involving quantum nonlinear
oscillator driven by an external field \cite{K90,LT94,LDT97,L96a}.
For this class of systems we are able to get the quantum states
that are very close for instance, to the FD coherent states
\cite{BWKL92,MPT94} or to the FD squeezed vacuum \cite{MLT98}.

\section{~~~II.~ FD coherent states generated by nonlinear
oscillator systems} \label{section1}

This section is devoted to the method of generation of the FD
coherent states  making a class of states defined in FD Hilbert
space. We shall concentrate on the states proposed by Bu\v{z}ek
et al. \cite{BWKL92} and further discussed by Miranowicz et al.
\cite{MPT94,MPOT95}, where both the Glauber displacement operator
and the states are defined in the FD Hilbert space \cite{MLI00}.
The method of generation discussed here is based on the quantum
systems containing a {\em \inx{Kerr medium}} represented by
nonlinear oscillator. It was introduced in Ref. \cite{LT94} as a
way of generating one-photon Fock states and was further adapted
for the FD coherent-state generation \cite{L97}. The model
discussed here represents a quantum nonlinear oscillator that
interacts with an external field. Systems of this kind can be a
source of various quantum states. For example, quantum nonlinear
evolution can lead to generation of squeezed states \cite{YIM86},
minimum uncertainty states \cite{KY86}, $n$-photon Fock states
\cite{K90,LT94,LDT97,L96a}, displaced Kerr states \cite{WBK91},
macroscopically distinguishable superpositions of two states
(Schr\"odinger cats) \cite{YS86,TM87} or higher number of states
(Schr\"odinger kittens) \cite{MTK90}. Of course, the problem of
practical realization of the system arises. At this point one
should emphasize that the most commonly proposed practical
realization is that in which a nonlinear medium is located inside
one arm of the Mach--Zehnder interferometer \cite{KY86}. However,
models comprising a quantum nonlinear oscillator can be achieved
in various ways. For instance, systems comprising trapped ions
\cite{BDKAK97}, trapped atoms \cite{WV97} or cavities with moving
mirror \cite{BJK97} can be utilized to generate states of our
interest.

\subsection{A.~~ Two-dimensional coherent states}
\inxx{coherent states; two-dimensional}

Let us start the discussion of practical possibilities of the FD
coherent-state generation from the simplest case, where only
superpositions of vacuum and single-photon state are involved (the
Hilbert space discussed is reduced to two dimensions). We consider
the system governed by the following Hamiltonian defined in the
interaction picture (in units of $\hbar=1$) to be
\begin{equation}
\label{N01}
\hat{H}(t)=\frac{\chi}{2}({\hat{a}}^\dagger)^2\hat{a}^2+
\epsilon({\hat{a}}^\dagger+\hat{a})f(t)
\end{equation}
where $\chi$ denotes the nonlinearity constant, which can be
related to the third-order susceptibility of the Kerr medium;
$\epsilon$ is the strength of the interaction with the external
field, and $\hat{a}^\dagger$ and $\hat{a}$ are bosonic creation
and annihilation operators, respectively. Moreover, using function
$f(t)$ we are able to define the shape of the envelope of external
field. For simplicity, we shall assume that the excitation is of
the constant amplitude and hence, we put $f(t)=1$. Obviously, one
should keep in mind that models discussed here concern a real
physical situation (although they naturally involve certain
limitation) and all operators, appearing in Eq. (\ref{N01}), are
defined in the infinite-dimensional Hilbert space.

Let us express the wavefunction for our system in the Fock basis
as
\begin{equation}
\label{N02} \left|\psi(t)\right>=\sum^{\infty}_{n=0}
C_n(t)\left|n\right>
\end{equation}
where the complex probability amplitude $C_n(t)$ corresponds to
the $n$th Fock state $\left|j\right>$ and determines its time
evolution. This wavefunction obeys the following Schr\"odinger
equation
\begin{equation}
\label{N03} i\frac{d}{dt}\left|\psi(t)\right>=\left(
\frac{\chi}{2}({\hat{a}}^\dagger)^2\hat{a}^2+
\epsilon({\hat{a}}^\dagger+\hat{a}) \right) \left|\psi(t)\right>
\end{equation}
for the Hamiltonian (\ref{N01}). Applying the standard procedure
to our wavefunction (\ref{N02}) and  Hamiltonian (\ref{N01}), we
obtain a set of equations for the probability amplitudes $C_n(t)$.
They are of the form
\begin{equation}
\label{N04} i\frac{d}{dt}C_n(t)=\frac{\chi}{2}n(n-1)C_n(t)
+\epsilon\left[ \sqrt{n}C_{n-1}(t)+\sqrt{n+1}C_{n+1}(t)\right]
\end{equation}
where $n$ corresponds to the $n$-photon Fock state. Obviously, one
should keep in mind that we deal with the infinite-dimensional
Hilbert space and so the set of equations for $C_n(t)$, given by
(\ref{N04}), is infinite too. However, our aim here is to show
that under special conditions our system behaves as one defined in
the FD Hilbert space. The first step is to assume that the
external excitation is weak ($\epsilon\ll\chi$). As a consequence,
we assume a perturbative approach. Moreover, and this is the main
point of our considerations, the part of Hamiltonian (\ref{N01})
corresponding to the nonlinear evolution of the system
\begin{equation}
\label{N05} \hat{H}_{\rm
NL}=\frac{\chi}{2}({\hat{a}}^\dagger)^2\hat{a}^2
\end{equation}
produces degenerate states corresponding to $n=0$ and $n=1$. As we
take into account not only the first part of Hamiltonian
(\ref{N01}) but also the second part, we see that a resonance
arises between the interaction described by the latter and the
degenerate states generated by $\hat{H}_{\rm NL}$. This resonance
and the weak interaction lead to a situation when the system
dynamics becomes of the closed form and cuts some subspace of
states out of all the $n$-photon Fock states. As a consequence,
assuming that the dynamics of the physical process starts from
vacuum $\left|0\right>$, the evolution of the system is restricted
to the states $\left|0\right>$ and $\left|1\right>$ solely. This
situation resembles in some sense the problem of two degenerate
atomic levels coupled by a zero-frequency field, where this
resonant coupling selects, from the whole set of atomic levels,
only those of them that lead to a closed system dynamics. For the
case discussed here our system evolution corresponds to the
two-level atom problem, where the interaction with remaining
atomic states can be treated as a negligible perturbation
\cite{AE75}. Obviously, one should note that the character of the
resonances commonly discussed in various papers, where the cavity
field and the difference between the energies of the atomic levels
(or cavity frequencies) have identical values, is different than
that of those discussed here.

Thus, we write following equations of motion
\begin{eqnarray}
\label{N06}
i\frac{d}{dt}C_0(t)&=&\epsilon C_1(t)\nonumber\\
i\frac{d}{dt}C_1(t)&=&\epsilon\left[C_0(t)+ \sqrt{2}C_2(t)\right]
\\
i\frac{d}{dt}C_2(t)&=&\chi C_2 (t)+ \epsilon\left[
\sqrt{2}C_1(t)+\sqrt{3}C_3(t)\right]\nonumber\\
&\vdots& \nonumber
\end{eqnarray}
for the probability amplitudes corresponding to the system
discussed here. Since we have assumed $\epsilon \ll\chi$, the
Eqs. (\ref{N06}) indicate that the amplitude $C_n (t)$ rapidly
oscillates in comparison with the amplitudes $C_{N}(t)$ if
$n>{N}$. Hence, analogously to the description of driven atomic
systems within the \inx{rotating wave approximation} (RWA)
\cite{AE75}, we neglect the influence of the probability
amplitudes $C_n(t)$ for $n\ge 2$. Therefore, the dynamics of our
system can be described by the following set of two equations
\begin{eqnarray}
\label{N07}
i\frac{d}{dt}C_0(t)&=&\epsilon C_1(t)\nonumber\\
i\frac{d}{dt}C_1(t)&=&\epsilon C_0(t)
\end{eqnarray}
and their solution
\begin{eqnarray}
\label{N08}
C_0(t)&=&i\cos (\epsilon t)\nonumber\\
C_1(t)&=&\sin(\epsilon t)
\end{eqnarray}
where we have assumed that the system starts its evolution from
vacuum $\left| 0\right>$. Clearly, this result resembles that for
a two-level atom in an external field \cite{AE75} and the dynamics
of the system exhibits well-known oscillatory behavior. This
result is identical to that derived for the simplest case (i.e.,
for $N=s+1=2$) of the FD generalized coherent states discussed by
us in the first part of this work \cite{MLI00}. Of course, one
should keep in mind that the set of Eqs. (\ref{N07}) gives
zero-order solutions in perturbative treatment. As a consequence,
the FD coherent states can be produced by the system discussed
within the error following from this approximation.

The preceding result concerns the situation where the external
excitation is characterized by a constant envelope: $f(t)=1$. For
the general case, the solution can be obtained easily, applying
the same procedure as for a resonantly driven two-level atom
\cite{AE75}. Then, the general solution can be expressed as
\begin{eqnarray}
\label{N09}
C_0(t)&=&i\cos \Theta (t)\nonumber\\
C_1(t)&=&\sin\Theta (t)
\end{eqnarray}
where the symbol $\Theta (t)$ denotes the pulse area and is
defined to be
\begin{equation}
\label{N10} \Theta(t)=\epsilon\int^t_0 f(t')dt'
\end{equation}

\subsection{B.~~ ${N}$-Dimensional coherent states}

It is possible to extend our considerations to the case of the FD
Hilbert space with arbitrary dimension. Similarly as in \cite{L97}
we introduce a system comprising a nonlinear oscillator with the
$N$th-order nonlinearity and governed by the following Hamiltonian
\begin{equation}
\label{N11}
\hat{H}(t)=\frac{\chi}{N}({\hat{a}}^\dagger)^{N}\hat{a}^{N}+
\epsilon({\hat{a}}^\dagger+\hat{a})f(t)
\end{equation}
The first term in (\ref{N11}) is the $N$-photon Kerr Hamiltonian
\cite{G87}, giving rise to optical bistability, and $\chi$ is
related to the ($2{N}-1$)-order susceptibility of the medium. The
second term in (\ref{N11}) represents coherent pumping modulated
by classical function $f(t)$. Similarly, as in the previous
section, we assume that the excitation has a constant envelope:
$f(t)=1$. Applying the procedure analogous to that described in
the previous section we get the following equations
\begin{eqnarray}
\label{N12}
i\frac{d}{dt}C_0(t)&=&\epsilon C_1(t)\nonumber\\
i\frac{d}{dt}C_1(t)&=&\epsilon\left[C_0(t)+
\sqrt{2}C_2(t)\right]\nonumber\\
&\vdots& \nonumber\\
i\frac{d}{dt} C_{{N}-1}(t)&=&\epsilon \left[\sqrt{{N}-1}
C_{{N}-2}(t)+
\sqrt{N}C_{N}(t)\right]\\
i\frac{d}{dt}C_{N}(t)&=&\chi ({N}-1)! C_{N}(t) +
\epsilon\left[\sqrt{N}C_{{N}-1}(t)+
\sqrt{{N}+1}C_{{N}+1}(t)\right]\nonumber\\
&\vdots& \nonumber
\end{eqnarray}
for the probability amplitudes $C_n(t)$. As it is assumed that
$\epsilon\ll\chi$, we can exclude all probability amplitudes
$C_n(t)$ for $n>{N}-1$. Hence, we get the set of equations in the
closed form and the dynamics of the system is practically
restricted within a space spanned over ${N}$ Fock states. For
instance, for ${N}=3$ Eqs. (\ref{N12}) reduce to
\begin{eqnarray}
\label{N13}
i\frac{d}{dt}C_0(t)&=&\epsilon C_1(t)\nonumber\\
i\frac{d}{dt}C_1(t)&=&\epsilon\left[C_0(t)+
\sqrt{2}C_2(t)\right] \\
i\frac{d}{dt}C_2(t)&=&\epsilon \sqrt{2}C_1(t)\nonumber
\end{eqnarray}
and have the solutions
\begin{eqnarray}
\label{N14} C_0(t)&=&\frac{1}{3}\left[2+\cos
\left(\sqrt{3}\epsilon t
\right) \right] \nonumber\\
C_1(t)&=&\frac{-i}{\sqrt{3}} \sin\left(\sqrt{3}\epsilon t
\right)\\
C_2(t)&=&\frac{\sqrt{2}}{3}\left[\cos\left(\sqrt{3}\epsilon t
\right) -1 \right]\nonumber
\end{eqnarray}
Again, these solutions are identical to those derived by
Miranowicz et al. \cite{MPT94} (compare Eq. (25) in Ref.
\cite{MLI00}). Of course, we can write the equations for arbitrary
value of the parameter ${N}$ and hence, get the formulas for the
probability amplitudes for the $n$-photon state expansion of the
FD coherent state defined in the ${N}$-dimensional Hilbert space.
In general, for any dimension ${N}$ and arbitrary real periodic
function $f(t)$ with the period $T$, we find that the system
evolves at $t=kT$ into the state \cite{MLDT96}
\begin{eqnarray}
\label{N15} |\phi(kT)\rangle &=& \sum_{n=0}^{{N}-1} C_n|n\rangle +
\epsilon C_{N} |{N}\rangle + {\cal O}(\epsilon^2)
\end{eqnarray}
where the superposition coefficients $C_n=\langle
n|\phi(kT)\rangle$ for $n=0,\dots,{N}-1$ are given by
\begin{equation}
\label{N16}
C_n=\frac{({N}-1)!}{N}\,\frac{(-1)^n}{\sqrt{n!}}\sum_{m=0}^{{N}-1}
\exp\left({i}k x_m  \epsilon c_0\right)\, \frac{{\rm
He}_n(x_m)}{[{\rm He}_{{N}-1}(x_m)]^2}
\end{equation}
and for $n={N}$ are
\begin{equation}
\label{N17} C_{N}=\sqrt{N} B C_{{N}-1} =(-1)^{{N}-1} B
\sqrt{\frac{({N}-1)!}{N}}\sum_{m=0}^{{N}-1} \frac{\exp(ik x_m c_0
\epsilon)}{{\rm He}_{{N}-1}(x_m)}
\end{equation}
Here, $x_m\equiv x_m^{({N})}$ are the roots of the \inx{Hermite
polynomial} of order ${N}$, ${\rm He}_{N}(x_m)=0$.  The
coefficient $B$ is defined to be
\begin{eqnarray}
\label{N18} B &=&
\frac{1}{2\pi}\sum_{n=-\infty}^{\infty}\frac{c_n}{n+a}
\end{eqnarray}
where
\begin{eqnarray}
\label{N19}
c_n=\int\limits_{0}^{T} f(t) \exp\left(-i2\pi
n\frac{t}{T}\right){\rm d}t
\end{eqnarray}
is the Fourier transform and $a=T\chi(N-1)!/(2\pi)$. In the first
part of this work (see Eq. (20) in Ref. \cite{MLI00}), we have
defined the $N$-dimensional generalized coherent states to be
($N\equiv s+1$) \inxx{coherent states; generalized}
\begin{eqnarray}
\label{N20} |\alpha\rangle_{(s)}  &=&
\exp\big[\alpha\hat{a}_s^{\dagger} -\alpha^* \hat{a}_s\big]
|0\rangle
\end{eqnarray}
in terms of the FD annihilation and creation operators,
\begin{eqnarray}
\hat{a}_s=\sum_{n=1}^s\sqrt{n} |n-1\rangle \langle n|, \qquad
\hat{a}_s^{\dagger}=\sum_{n=1}^s\sqrt{n}|n\rangle \langle n-1|
\label{N21}
\end{eqnarray}
respectively. On omitting terms proportional to $\epsilon$, we
explicitly show that
\begin{eqnarray}
\label{N22} |\alpha=-ikc_0\epsilon\rangle_{(s)} &=&
|\phi(kT)\rangle + {\cal O}(\epsilon)
\end{eqnarray}
Thus, the state created in the process governed by the Hamiltonian
(\ref{N11}) is the finite-dimensional coherent state.

\section{~III.~ Numerical calculations}

It is possible to verify our considerations performing appropriate
numerical calculations. As, we have excluded here all damping
processes, the dynamics of our system can be described by the
unitary evolution. Therefore, we define the unitary evolution
operator
\begin{equation}
\label{N23}
\hat{U}=\exp\left\{-i\left[\frac{\chi}{N}(\hat{a}^\dagger)^{N}
\hat{a}^{N} +\epsilon (\hat{a}^\dagger +\hat{a})\right]t\right\}
\end{equation}
on the basis of Hamiltonian (\ref{N01}). In (\ref{N23}) all
operators are defined in the ${N}$-dimensional Hilbert space. For
example, for ${N}=4$ the wavefunction $\left|\psi(t) \right>$ can
be expressed as
\begin{eqnarray}
\label{N24}
\left| \psi (t)\right> &=&\left(
\begin{array}{c}
C_0(t)\\
C_1(t)\\
C_2(t)\\
C_3(t)
\end{array}
 \right)
 \end{eqnarray}
whereas the annihilation and creation operators ($\hat{a}$ and
$\hat{a}^\dagger$ respectively) can be represented by the
following matrices
\begin{eqnarray}
\label{N25} \hat{a}=\left[
\begin{array}{llll}
0 & 1 & 0 & 0\\
0 & 0 & \sqrt{2} & 0\\
0 & 0 & 0 & \sqrt{3}\\
0 & 0 & 0 & 0
\end{array}
\right], \qquad \hat{a}^\dagger =\left[
\begin{array}{llll}
0 & 0 & 0 & 0\\
1 & 0 & 0 & 0\\
0 &  \sqrt{2} & 0 & 0\\
0 & 0 & \sqrt{3} & 0
\end{array}
\right]
\end{eqnarray}
which are special cases of (\ref{N21}) for $s=3$. As a
consequence, the Hamiltonian (\ref{N11}) can be constructed using
the Eq. (\ref{N25}) matrix representations. Next we should
construct the evolution operator $\hat{U}$. Since this operator
is in the form of the matrix exponential it could be necessary to
solve eigensystem with the Hamiltonian $\hat{H}$. This step can be
easily done by applying standard numerical procedures
\cite{PFTV86}. Obviously, other methods of calculating matrix
exponentials can be utilized as well. For instance, the
Taylor-series expansion of the operator $\hat{U}$ can be helpful
in this case. Using the evolution operator derived, we are in a
position to generate the wavefunction for arbitrary time $t$.

Thus, assuming that the system starts its evolution from vacuum
$\left|0\right>$ we act (numerically) $\hat{U}$ on the wave
function of the system represented by the ${N}$-element vector
\begin{eqnarray}
\label{N26}
\left|\psi (0)\right> &=&\left(
\begin{array}{c}
1\\ 0\\ 0\\ \vdots\\ 0
\end{array}
 \right)
\end{eqnarray}
and obtain the vector representation of the desired wavefunction
$\left|\psi (t)\right>$ corresponding to the state of our system
for the time $t$:
\begin{equation}
\label{N27} \left|\psi (t)\right>=\hat{U}\left|\psi (0)\right>
\end{equation}
It would be interesting to compare Eqs. (\ref{N26}) and
(\ref{N27}) with the Glauber definition of the coherent state
\cite{G63}
\begin{equation}
\label{N28} \left|\alpha\right>_{(\infty)} =\hat{D}(\alpha
,\alpha^*)\left| 0\right>
\end{equation}
where the Glauber \inx{displacement operator} $\hat{D}(\alpha
,\alpha^*)$ is defined as
\begin{equation}
\label{N29} \hat{D}(\alpha
,\alpha^*)=\exp\left(\alpha\hat{a}^\dagger -\alpha^*\hat{a}\right)
\end{equation}
It is seen that the operator $\hat{U}$ defined in Eq. (\ref{N23})
plays the same role as the Glauber displacement operator
$\hat{D}(\alpha ,\alpha^*)$. Obviously, it should be kept in mind
that $\hat{U}$ is defined in the FD Hilbert space, contrary to the
definition of $\hat{D}$ in which the space has been assumed to be
infinite dimensional. Therefore, we conclude that within the
\linebreak
\begin{figure}[ht]
\vspace{-3mm}
\centerline{\psfig{figure=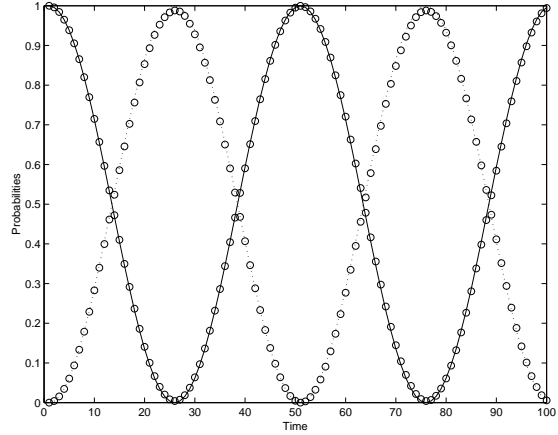,height=6.0cm}}
\caption{Time-evolution of the probabilities (analytical results)
for vacuum $\left| 0\right>$ (solid curve) and one-photon state
$\left| 1\right>$ (dotted curve) for the system with Kerr medium
described by the Hamiltonian $\frac{1}{2} \chi(\hat{a}^\dagger)^2
\hat{a}^2$. The circle marks denote numerical results. The pulse
strength is $\epsilon =\pi /50$.\label{leonfg01}} \vspace{-5mm}
\end{figure}
\begin{figure}[ht]
\centerline{\psfig{figure=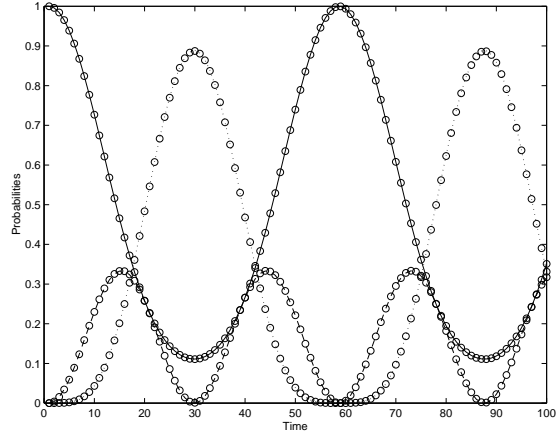,height=6.0cm}} \caption{The
same as in Fig. \ref{leonfg01} but for the system with the Kerr
medium described by $\frac{1}{3}\chi(\hat{a}^\dagger)^3
\hat{a}^3$. Analytical results for: vacuum $\left| 0\right>$
(solid curve), one-photon state $\left| 1\right>$ (dashed curve),
and two-photon state $\left| 2\right>$ (dotted curve). The
numerical results are marked by circles.\label{leonfg02}}
\end{figure}
\noindent assumptions introduced here we deal with the following
correspondence:
\begin{equation}
\label{N30} \left.\hat{U}\right|_\infty \leftrightarrow \hat{D}
\end{equation}
To check our analytical formulas derived in the previous sections,
we shall concentrate on the two cases where the parameter
${N}=2,3$. First, for ${N}=2$ we apply the evolution operator
\begin{equation}
\label{N31}
\hat{U}=\exp\left\{-i\left[\frac{\chi}{2}\left(\hat{a}^\dagger\right)^2
\hat{a}^2 +\epsilon \left(\hat{a}^\dagger
+\hat{a}\right)\right]t\right\}
\end{equation}
and the results are shown in Fig. \ref{leonfg01}, which also shows
the analytical results for the probabilities of finding the system
in vacuum $\left| 0\right>$ and one-photon $\left| 1\right>$
states together with those of the numerical method. The analytical
and numerical results agree almost perfectly and for these two
cases we obtain the well-known oscillatory behavior. Obviously,
one should keep in mind that the interaction with the external
field is weak ($\epsilon\ll\chi$) and we assume that
$\epsilon=\pi/50$ contrary to $\chi=1$.

Analogously, for ${N}=3$ three states are involved in the system
evolution. For this case the evolution operator should be of the
form
\begin{equation}
\label{N32}
\hat{U}=\exp\left(-i\left[\frac{\chi}{3}\left(\hat{a}^\dagger\right)^3
\hat{a}^3 +\epsilon \left(\hat{a}^\dagger
+\hat{a}\right)\right]t\right)
\end{equation}
whereas the parameters $\epsilon$ and $\chi$ are the same as for
the case of ${N}=2$. Similarly as for ${N}=2$ the agreement of the
analytical results with their numerical counterparts is very good.
Thus, Fig. \ref{leonfg02} depicts oscillations of the
probabilities for the states $\left| 0\right>$, $\left| 1\right>$
and $\left| 2\right>$. The amplitude of the oscillations for
one-photon state $\left| 1\right>$ is considerably smaller than
that for other two states involved in the evolution. This fact
agrees with the properties of the Fock expansion of the FD
coherent state \cite{MPT94}.

\begin{figure}[ht]
\centerline{\psfig{figure=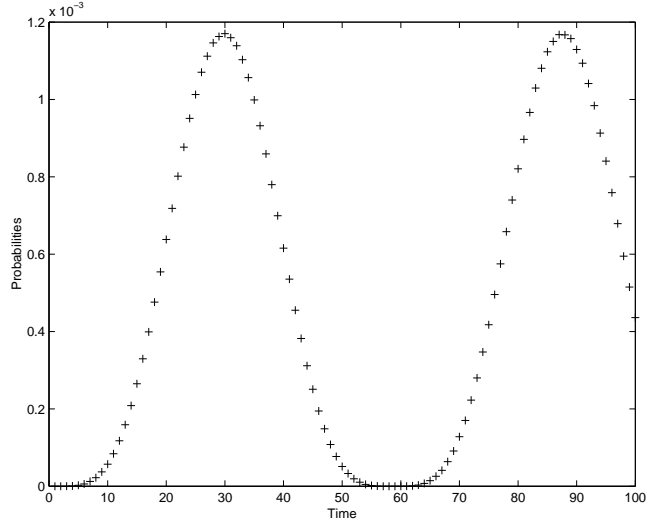,height=7.0cm}} \caption{The
probability for the three-photon state $\left| 3\right>$ obtained
from the numerical calculations. All parameters are the same as
for Fig. \ref{leonfg02}.\label{leonfg03}} \vspace{-5mm}
\end{figure}

Applying the numerical method described here, we can also estimate
the error of the perturbative treatment introduced in the previous
sections. In Fig. \ref{leonfg03} we show the probability
corresponding to the three-photon state $\left| 3\right>$ as a
function of time. It is seen that the probability oscillates in a
similar way as those corresponding to the states $\left|
0\right>$, $\left|1\right>$, and $\left| 2\right>$. However, the
amplitudes of the oscillations differ significantly. Thus, the
probability for the state $\left| 3\right>$ oscillates between $0$
and $\sim 1.2\times 10^{-3}$ whereas that corresponding to the
state $\left| 1\right>$ changes its value from $0$ to $\sim 0.3$
(Fig. \ref{leonfg02}). We see that the dynamics of the system
described by the Hamiltonian (\ref{N01}) is restricted in practice
to the closed set of the Fock states. This fact and the behavior
of the probabilities shown in Figs. \ref{leonfg01} and
\ref{leonfg02} proves that the quantum states generated by the
system described by Hamiltonian (\ref{N01}) are very close to the
FD coherent states described in Ref. \cite{MPT94}.

\section{~IV.~ State generation in dissipative systems}
\inxx{dissipative systems}

It is obvious that in the real physical situations we are not able
to avoid dissipation processes. For dissipative systems, we cannot
take an external excitation too weak (the parameter $\epsilon$
cannot be too small) since the field interacting with the
nonlinear oscillator could be completely damped and hence, our
model could become completely unrealistic. Moreover, the
dissipation in the system leads to a mixture of the quantum states
instead of their coherent superpositions.Therefore, we should
determine the influence of the damping processes on the systems
discussed here. To investigate such processes we can utilize
various methods. For instance, the quantum jumps simulations
\cite{DCM92} and quantum state diffusion method \cite{GP92} can be
used. Description of these two methods can be found in Ref.
\cite{GK94} where they were discussed and compared. Another way to
investigate the damping processes is to apply the approach based
on the density matrix formalism. Here, we shall concentrate on
this method \cite{MH86,MH91,LT94}.

As we have discussed earlier, the time dependence of the envelope
of external excitation does not influence the final analytical
result discussed here. The parameter $f(t)$ appears only inside
the integral determining the external pulse area $\theta(t)$ [Eq.
\ref{N10}]. Therefore, we can assume without losing generality of
our considerations that the excitation is in the form of a series
of ultrashort pulses. Then the function $f(t)$ can be modeled by
the Dirac-delta functions as
\begin{eqnarray}
\label{N33} f(t)&=&\sum_{k=0}^{\infty} \delta (t-k T)
\end{eqnarray}
where $T$ is a time between two subsequent pulses. For such
situation the time-evolution of the system can be divided into two
different stages. When the damping processes are absent, the first
stage is a ``free'' evolution of the nonlinear oscillator
determined by the unitary evolution operator
\begin{eqnarray}
\label{N34} \hat{U}_{\rm NL}&=&\exp \left[-i\frac{\chi T}{2}\left(
\hat{a}^\dagger\right)^2\hat{a}^2 \right]
\end{eqnarray}
We assume the simplest case, where the time-evolution is
restricted to two quantum states $\left| 0\right>$ and $\left|
1\right>$. The second stage of the time-evolution of the system is
caused by its interaction with an infinitely short external pulse.
This part of the evolution is described by the second term of the
Hamiltonian (\ref{N01}) and can be described by the following
evolution operator
\begin{eqnarray}
\label{N35} \hat{U}_{\rm
K}&=&\exp\left[-i\epsilon\left(\hat{a}^\dagger
+\hat{a}\right)\right]
\end{eqnarray}
The overall evolution of the system can be described as a
subsequent action of the operators $\hat{U}_{\rm NL}$ and
$\hat{U}_{\rm K}$ on the initial state. When we take into account
losses during the time-evolution between two pulses we should
solve the appropriate master equation. It can be written as
\begin{eqnarray}
\label{N36} \frac{d\rho}{dt}= -i\frac{\chi}{2}
\left(\hat{a}^\dagger\right)^2\hat{a}^2
+\frac{\gamma}{2}\left(2\hat{a}\rho\hat{a}^\dagger
-\hat{a}^\dagger \hat{a}\rho-\rho\hat{a}^\dagger \hat{a}\right)
\end{eqnarray}
The solution of this master equation in the Fock number states
basis is given by \cite{MH86,MH91}
\begin{eqnarray}
\label{N37} \left<p\right|\rho(t+T)\left|q\right>=
\exp\left[i\frac{\vartheta}{2}(p-q)\right]
\frac{[g(T)]^{(p+q)/2}}{\sqrt{p!q!}}
\sum_{n=p}^{\infty}\left<n\right|\rho
(t)\left|n-(p-q)\right> \nonumber \\
\times \frac{\sqrt{n![n-(p-q)]!}}{(n-p)!}(1+i\delta
)^{-(n-p)}\left[1-g(T)\right]^{(n-p)}\quad\quad
\end{eqnarray}
where
\begin{eqnarray}
\label{N38} \delta &=&\frac{p-q}{\kappa}
\nonumber \\
g(T)&=&\exp \left[-\kappa\vartheta-i\vartheta(p-q)\right]
\nonumber \\
\kappa&=&\frac{\gamma}{\chi}
\nonumber \\
\vartheta&=&\chi T
\end{eqnarray}
The symbol $\gamma$ appearing above is a damping constant
responsible for the cavity loss. Thus solving the master equation
(\ref{N36}) we can determine the probabilities of finding the
system in an arbitrary $n$-photon state. Of course, the evolution
during single ultrashort, external pulse is determined by the
operator $\hat{U}_K$ as before.
\begin{figure}[ht]
\vspace*{-2.5cm} \hspace*{5mm}
\centerline{\psfig{figure=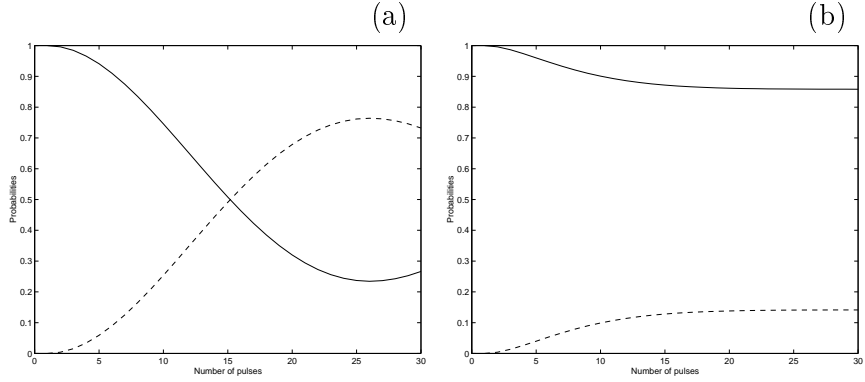,height=24cm}}
\vspace*{-16.8cm} \caption{The probabilities for vacuum $\left|
0\right>$ (solid curve) and one-photon state $\left| 1\right>$
(dashed curve) for the Kerr medium described by $\frac
{1}{2}\chi(\hat{a}^\dagger)^2 \hat{a}^2$. The damping constants
are: (a) $\gamma =0.01$ and (b) $\gamma =0.1$. The pulse strength
is $\epsilon =\pi /50$ and the time $T=\pi$.\label{leonfg04}}
\end{figure}

Thus Fig. \ref{leonfg04} shows probabilities for vacuum $\left|
0\right>$ and one-photon $\left| 1\right>$ state for weak external
excitation once more. We have chosen two values of the damping
parameter: $\gamma=0.1$ (Fig. \ref{leonfg04}(a)) and $\gamma=0.01$
(Fig. \ref{leonfg04}(b)). It is seen that for weak damping we
observe slow oscillations of the probabilities, similarly as for
the case of the quantum nonlinear oscillator without dissipation.
Moreover, for ($\gamma =0.01$) the amplitude of the oscillations
reaches over $75\%$ of its value for the case of $\gamma =0$. As a
consequence, we are able to get the field very close to the
desired quantum state. However, as the damping increases the
situation changes considerably. For $\gamma =0.1$ the oscillations
of the probabilities vanish, and the resulting state is far from
the FD coherent state defined in the two-dimensional Hilbert
space. We see that the dissipation in the system can drastically
lower the effectiveness of producing the FD coherent states.
Nevertheless, one should keep in mind one of the crucial points
of our considerations: the assumption of weak external excitation.
Hence, we hope that for sufficiently weak damping, our system can
evolve to a state that is very close the quantum state of our
interest.

\section{~~V.~ Generalized method for FD squeezed vacuum
generation} \inxx{squeezed vacuum; generalized}

The method described in the previous sections can be easily
generalized to be useful for generation of various FD
quantum-optical states different from the FD coherent state. Thus,
we shall show an example of how to adapt our method to generate
the FD squeezed vacuum \cite{MLT98}. In the first part of this
work [see Eq. (78) in Ref. \cite{MLI00}], we have defined the
$(s+1)$-dimensional generalized squeezed vacuum to be
\begin{eqnarray}
\label{N39} \left|\xi\right>_{(s)}=\exp\left[\frac{\xi}{2}
(\hat{a}_s^{\dagger})^2-\frac{\xi
^{*}}{2}\hat{a}_s^{2}\right]|0\rangle
\end{eqnarray}
where $\xi=|\xi |\exp (i\phi)$ is the complex squeeze parameter,
whereas $\hat{a}_s$ and $\hat{a}_s^{\dagger}$ are, respectively,
the FD annihilation and creation operators defined by (\ref{N21}).
Since the properties of the FD squeezed vacuum have already been
discussed \cite{MLI00}, here we shall concentrate on the method
of its generation.

We assume that our system consists of a Kerr medium of the
$(s+1)$th-order nonlinearity and a parametric amplifier driven by
a series of ultrashort external classical-light pulses. Thus, the
Hamiltonian describing our system can be written in the
interaction picture as
\begin{eqnarray}
\label{N40} \hat{H}=\frac{\chi}{(s+1)}
\left(\hat{a}^\dagger\right)^{s+1}\hat{a}^{s+1}
+\epsilon\left(\hat{a}^{\dagger 2}+\hat{a}^{2}\right)f(t)
\end{eqnarray}
where the first term describes the $(s+1)$-photon nonlinear
oscillator (Kerr medium) as in (\ref{N11}), and the second term
represents a pulsed parametric oscillator modulated by $f(t)$,
given by (\ref{N33}). This situation differs from those discussed
in the previous sections in one important point, namely, in the
character of external excitation. For this case we assume that the
oscillator is driven by a second-order parametric process instead
of linear excitation involved in the FD coherent-state generation.
The model, described by (\ref{N40}) with the two-photon Kerr
Hamiltonian, was studied by Milburn and Holmes \cite{MH91} in
their analysis of quantum coherence and classical chaos. The
system for $s=1$ and $f(t)=1$ is referred to as the {\em
\inx{Cassinian oscillator}} and has been analyzed in the context
of squeezing by, for instance, Gerry et al. \cite{GR87} and
DiFilippo et al. \cite{DNBP92}. The time evolution of the system
leads to the generation of the quantum states that differ
significantly from the FD coherent states. In a similar way as
for the generation of the latter, we assume that the excitation is
weak ($\epsilon\ll \chi$), and we can apply the perturbative
treatment again. As a consequence, we get the formula for the
$n$-photon state expansion
\begin{eqnarray}
\label{N41} \left|\phi (t)\right>=\sum_{n=0}^{\sigma}C_{2n}
(t)\left| 2n\right> +\epsilon C_{2\sigma +2} (t) \left| 2\sigma
+2\right> +O(\epsilon ^2)
\end{eqnarray}
where the expansion coefficients $C_{2n}=\left< 2n\right|\left.
\phi (t)\right>$ for $n=0,\dots ,\sigma$ are
\begin{eqnarray}
\label{N42} C_{2n} (t)=(-1)^n \frac{(2\sigma)!}{\sqrt{(2n)!}}
\sum_{k=0}^{\sigma} \exp (i x_k \epsilon t) \frac {G_n
(x_k)}{G_\sigma (x_k)G'_{\sigma +1} (x_k)}
\end{eqnarray}
and
\begin{eqnarray}
\label{N43} C_{2\sigma +2}(t)=2^{-\sigma -1}\sqrt{(2\sigma
+1)(2\sigma +2)}\, C_{2\sigma}(t)
\end{eqnarray}
The functions $G_n (x)$ appearing above are the
\inx{Meixner--Sheffer orthogonal polynomials}; the prime sign in
Eq. (\ref{N42}) denotes their $x$-derivative,  and $\sigma = {\rm
Int}(s/2)$ is the integer part of $s/2$. If we omit the terms
proportional to $\epsilon$ and higher, we get the expansion for
the FD squeezed vacuum as
\begin{eqnarray}
\label{N44} \left|\xi =-2\epsilon t\right>_{(s)}=\left|\phi
(t)\right> + O(\epsilon)
\end{eqnarray}
So, our system evolves to a state close to the FD squeezed vacuum
discussed in Ref. \cite{MLT98}.

\section{VI.~ Summary}

We have discussed one of the possible methods of generation of the
FD quantum-optical states. Although, it is possible to generate
$n$-photon Fock states and then to construct a desired state from
these states, we have concentrated on the generation schemes that
can lead directly to the FD coherent states and FD squeezed
vacuum. The method described here is based on the quantum
nonlinear oscillator evolution. We have assumed that this
oscillator is driven by an external excitation. We have shown that
within the weak excitation regime we are able to generate  with
high accuracy the appropriate FD quantum state. Thus, depending on
the character of the excitation we can produce various FD states.
For instance, for the linear excitation case we generate the FD
coherent state, whereas for the parametric excitation of the FD
squeezed vacuum can be achieved. Moreover, we have shown that the
mechanism of the generation does not depend on the shape of the
excitation envelope. Hence, various forms of the latter can be
assumed depending of the feasibility of our model from the
experimental or mathematical point of view.

For the situations discussed here appropriate analytical formulas
for the generated states have been derived. These results have
been obtained within the perturbation theory, and they agree with
those of the $n$-photon expansion of the appropriate FD states.
Moreover, we have proposed methods for checking our results
numerically, and we have shown that numerical results agree very
well with the analytical ones. Since, we are not able to avoid
dissipation processes from real physical situations, we have
discussed damping processes two. It has been shown that although
dissipation can play crucial role in the whole system dynamics
and is able to destroy the effect of the FD state generation
completely, under special assumptions these states can be
achieved.

\subsection*{ACKNOWLEDGMENTS}

The authors thank J. Bajer, S. Dyrting, N. Imoto, M. Koashi, T.
Opatrn\'y, \c{S}. K. \"Ozdemir, J. Pe\v{r}ina, K. Pi\c{a}tek and
R. Tana\'s for their helpful discussions. A. M. is indebted to
Prof. Nobuyuki Imoto for his hospitality and stimulating research
at SOKEN.



\printindex

\end{document}